\documentclass[preprint]{aastex62}

\usepackage{natbib}
\usepackage{graphicx, epsfig, hyperref}
\def\up#1{\leavevmode \raise.16ex\hbox{#1}}

\newcommand\kms{${\rm km\,s}^{-1}$}
\newcommand\kmss{${\rm km\,s}^{-1}$~}

\newcommand\teffs{$T_{\rm eff}$~}
\newcommand\teff{$T_{\rm eff}$}

\received{August 27th 2020}
\revised{September 15th 2020}
\accepted{October 13th 2020}

\submitjournal{ApJ}

\shorttitle{The Photospheric Temperatures of Betelgeuse During the Great  Dimming}
\shortauthors{Harper et al.}

\begin{document}

\title{The Photospheric Temperatures of Betelgeuse during the Great Dimming of 2019/2020:  No New Dust Required}

\correspondingauthor{Graham M. Harper}
\email{graham.harper@colorado.edu}

\author[0000-0002-7042-4541]{Graham M. Harper}
\affiliation{Center for Astrophysics and Space Astronomy\\
University of Colorado Boulder \\
389 UCB\\
Boulder, CO 80309\\
USA}

\author{Edward F. Guinan}
\affiliation{Astrophysics and Planetary Science Department\\
Villanova University\\
Villanova, PA 19085\\
USA}

\author{Richard Wasatonic}
\affil{Astrophysics and Planetary Science Department\\
Villanova University\\
Villanova, PA 19085\\
USA}

\author{Nils Ryde}
\affiliation{Lund Observatory\\
Department of Astronomy and Theoretical Physics\\
Lund University\\
Box 43, SE-221 00 Lund\\
Sweden}

\begin{abstract}

The processes that shape the extended atmospheres of red supergiants, heat their chromospheres, create molecular reservoirs, drive mass loss, and create dust remain poorly understood.  Betelgeuse's $V$-band "Great Dimming" event of 2019 September /2020 February and its subsequent rapid brightening  provides a rare opportunity to study these phenomena.  Two different explanations have emerged to explain the dimming; new dust appeared in our line of sight attenuating the photospheric light, or a large portion of the photosphere had cooled.  Here we present five years of Wing three-filter ($A$, $B$, and $C$ band) TiO and near-IR photometry obtained at the Wasatonic Observatory. These reveal that parts of the photosphere had a mean effective temperature
$(T_{\rm eff}$) significantly lower than that found by \citet{Levesque_2020ApJ...891L..37L}.  Synthetic photometry from MARCS -model photospheres and spectra
reveal that the $V$ band, TiO index, and $C$-band photometry, and previously reported
4000-6800\AA{} spectra can be quantitatively reproduced if there are multiple photospheric components, as hinted at by Very Large Telescope (VLT)-SPHERE images \cite{montarges_2020}.  If the cooler component has 
$\Delta T_{\rm eff} \ge 250$\,K cooler than 3650\,K, then no new dust is required to explain
the available empirical constraints.  A coincidence of the dominant short- ($\sim 430$\,day) and 
long-period ($\sim 5.8$\,yr) $V$-band variations occurred near the  time of deep minimum \citep{2019ATel13365....1G}. This is in tandem with the strong correlation of $V$ mag and photospheric radial velocities, recently reported by \citet{Dupree_2020ApJ...899...68D}. These suggest that the cooling of a large fraction of the visible star has a dynamic origin related to the photospheric motions,
perhaps arising from pulsation or large-scale convective motions. 

\end{abstract}

\keywords{stars: individual: $\alpha$~Ori -- stars: circumstellar matter -- stars: mass loss --lines: profiles}

\section{Introduction} \label{sec:intro}

 In 2020 February Betelgeuse ($\alpha$~Orionis; $\langle{V}\rangle\sim 0.45$; M1-2 Ia-Iab) became the faintest it has been on record in the visual/$V$ band ($V=+1.61$) --- the Great Dimming of 2019/2020. At this time the bright red supergiant (RSG) thus attracted much attention 
 with the public and in the amateur and professional astronomy communities.  Betelgeuse is a semi-regular variable that has been known to vary, at least in the modern era, since 1839 \citep{Herschel_1840MmRAS..11..269H}. It typically shows two dominant periods at $V$ band and in its photospheric radial velocity. Since 1985 the shorter period has been near 420
 days, while the longer period prior to 1933 was $\simeq 5.87$~years 
(\citep[][and references therein]{Goldberg_1984PASP...96..366G}, \citet{Dupree_1987ApJ...317L..85D} and \citet{Myron_1989AJ.....98.2233S}). Analysis of visual observations made by AAVSO observers from 1918 to 2006 \citep{Kiss_2006MNRAS.372.1721K} yields similar, but slightly different,
 results with periods of $388 \pm 30$\,days and $5.61 \pm 1.1$\,years, respectively.
 Both of these periods vary in length and amplitude\footnote{We estimate a short period of $430 \pm 10$\,day from our 25 years of multi-epoch photometry at $V$}.  The variations are typically $\Delta V \simeq 0.5$\,mag peak-to-peak but during the Great Dimming $\Delta V\simeq 1.1$\,mag and the fading was visible to the naked eye when viewing the constellation Orion.  Observations were made at many observatories,
e.g., \citet{Chandra_2020ATel13501....1K} reported a non-detection with the Chandra X-ray telescope, \citet{Harper_2020ApJ...893L..23H} reported that SOFIA-EXES observation showed the
circumstellar emission line-to-continuum flux ratio was unchanged with respect to the normal state, and \citet{Gehrz_2020ATel13518....1G} reported that multi-band infrared photometry
between 1.2 and 8.8$\mu$m was essentially unchanged from previous observations. No exotic high-energy phenomena or new dust emission was detected.

The origin of the dominant short and long periods observed on red supergiants (RSGs) is not clearly established. It has been proposed that the shorter period is a fundamental, or low-order overtone radial pulsation, e.g., \citet{Chatys_2019MNRAS.487.4832C}, \citet{Soraisam_2020ApJ...893...11S}, and
for Betelgeuse  \citet{Joyce_2020ApJ...902...63J}, conclude that the short period is a fundamental mode driven by the $\kappa$-mechanism.  
The long secondary period may be related to flow timescales of giant convection cells \citep{Stothers_2010ApJ...725.1170S}. Of particular interest for Betelgeuse is that the minimum of the relatively well-defined short period appeared to coincide with the long-period minimum
during 2020 February \citep{Guinan_2019ATel13341....1G}. In addition to causing the star to be dimmer overall  by about $0.4$\,mag, it is possible that this coincidence may have a bearing on the deep minimum observed, which is greater than a simple addition of amplitudes, perhaps reflecting a nonlinear response. 

VLT-SPHERE observations\footnote{https://www.eso.org/public/images/eso2003c/}, made 
with the H$\alpha$-continuum filter, in 2019 January and December show that the brightness across the southern photosphere had dimmed markedly in December (Montarg\`es et al. 2020).  The spatial scale of the dimmed region is compatible with large convection cells that
are expected and observed to be a significant fraction of the stellar radius \citep[e.g.,][]{Wilson_1997MNRAS.291..819W, Freytag_2002AN....323..213F,  Montarges_2016A&A...588A.130M, Lopez_2018A&A...620A.199L}.  These photospheric phenomena are expected to be related to the mechanisms that
ultimately heat RSG chromospheres, perhaps through acoustic or magnetic shocks, fuel the quasi-static molecular reservoirs that sit between the upper photosphere and chromosphere \citep{Eamon_2020A&A...638A..65O}, drive mass loss, and create dust, all of which are poorly understood processes. 

The $V$ band is dominated by molecular absorption from titanium oxide (TiO) molecules, and is sensitive
to the temperature of the stellar atmosphere.  The reduction in $V$-band brightness might be expected to correspond to a significant reduction in photospheric temperature across a large fraction of the 
visible photosphere. However,  \citet{Levesque_2020ApJ...891L..37L} used the strength of TiO absorption bands in the 
4000--6800\AA{} spectral region to derive an effective temperature $T_{\rm eff}=3600\pm 25$\,K for 2020 February 15. This is near the time when the star was faintest at $V=+1.61$\,mag. This is only a small decrease from  $T_{\rm eff}=3650 \pm 25$\,K when $V\simeq  0.5$
\citep{Levesque_2020ApJ...891L..37L}. They invoke
large dust grains to explain the large  decrease in $V$-band brightness.
\citet{JCMT_2020ApJ...897L...9D} reported that 450 and 850\,$\mu$m  sub-mm fluxes declined by $\sim 20$\% during the dimming period.  Spatially resolved ALMA observations show that the 870$\mu$m flux originates from the extended upper photosphere or  lower chromosphere, near $1.3R_\ast$ \citep{Eamon_2017A&A...602L..10O}. This wavelength region is on the Rayleigh-Jeans tail of the Planck function, indicating a reduction in the mean local gas temperature also by 20\% if the angular size of the region remained constant.
\citet{JCMT_2020ApJ...897L...9D} also point out that a large region of lower \teffs can qualitatively reproduce
their observations and the \citet{Levesque_2020ApJ...891L..37L}  spectra.  More recently \citet{Dupree_2020ApJ...899...68D} discussed results from spatially resolved Hubble Space Telescope near-ultraviolet (NUV) spectroscopy. They report on the enhancement of chromospheric Mg~II 2800\AA{} h \& k
emission during 2019 October, 3 to 4 months prior to the Great Dimming. Their study linked
the Mg~II emission to mass ejection, perhaps leading to dust formation. 

While Betelgeuse appeared close to the Sun, as seen from Earth, in 2020 June/July, the STEREO Solar Mission obtained images of that region of the sky
\citep{Dupree_2020ATel13901....1D}. The star decreased in brightness by 30\% from its 2020 late-April maximum, declining to $V \simeq +0.80$\,mag by the end of July.  Our photometry, along with photometry from D. Carona (2020, private communication), define a local light minimum in mid-August 
($V = +0.85$\,mag) and slowly rising to $V = +0.68$\,mag by early-October 2020.

{  The origin of the unprecedented large dimming of Betelgeuse during 2019/2020 is the subject of this paper.} Here we report Wing three-filter and  $V$-band photometry of Betelgeuse for the last 5 years, including good coverage of the Great Dimming event.  We analyze the data with synthetic photometry from MARCS models to shed light on the origin of the reduction in $V$-band brightness during 2019/2020.
 In \S2 the Wing three-filter near-IR observations and synthetic photometry are described, and the results are presented in \S3. The discussion of the findings is given in \S4, and conclusions are given in \S5.

\section{Observations}

Photometric observations of Betelgeuse have been conducted at Wasatonic Observatory (Allentown, PA) since 
1996 September.  The observations were carried out 
with a $f/10$ 20-cm (8 inch) Celestron Schmidt-Cassegrain Telescope (SCT) for the first five 
observing seasons, followed by using a $f/10$ 28-cm (11 inch) SCT for the remaining 19 seasons. 
An uncooled Optec SSP-3 photometer was employed. The SSP-3 has a red-near-IR sensitive 
PIN-photodiode detector with a broad spectral range from 3000 to 11000\AA.

$V$-band photometry was conducted using a wide-band V filter (5550\AA) that 
closely matches the $V$ bandpass of the Johnson $UBV$ system. In addition, three 
intermediate band Wing red/near-IR filters were employed that form a subset of Wing’s 
eight-color system \citep{WW1978ApJ...222..209W, Wing_1992JAVSO..21...42W, Wasatonic_2015PASP..127.1010W}.   The Wing $A$ filter is centered on the strong $\gamma(0, 0)$ TiO 7190\AA\ band head, while the central wavelengths of the Wing $B$ (7500\AA) 
and $C$ (10240\AA) filters are centered on near-IR continuum regions that are free 
of strong absorption lines in M stars,  see Table~\ref{tab:filters}.  Additional observing details can be found in the Appendix.

\begin{table}
\begin{center}
\caption{Characteristics of the Four Photometric Filters}
\label{tab:filters}
\begin{tabular}{lccc}
\tableline\tableline
Filter    &  Spectral Region & Central-$\lambda$ &  Bandpass (FWHM)  \\ 
            &                              &   (\AA)                               & (\AA) \\ \tableline
$V$           &                             &  5550    &  910 \\ 
$A $          &  TiO $\gamma(0,0)$   & 7190    & 115 \\ 
$B$          &   Continuum        & 7540            & 110 \\
$C$          &   Continuum       &  10240         & 445  \\
\tableline
\end{tabular}
\end{center}
\end{table}

\subsection{TiO Photometry}

Data from the first (1996/97) season were reported by \citet{Morgan_1997IBVS.4499....1M}, but
here we report on the most recent 5 years of photometry pertinent to the 2019/2020 Great Dimming. To estimate a typical nightly magnitude uncertainty for the Wing photometry, we note the strong linear correlation
between  5 years of $V$ and Wing $A$-band magnitudes (Pearson Linear Coeff. =0.973, $N=209$). This is not surprising given the presence of strong TiO absorption in both bands. Thus, for evolved M stars like Betelgeuse, changes in $V$ magnitude are more of a measure of \teffs than they are of stellar luminosity. Similar but slightly weaker correlations are found for $B$ and $C$ magnitudes. 

{\it Hipparcos} photometry \citep{ESA_1997ESASP1200.....E} shows that Betelgeuse varies smoothly on time-scales of days, while the observed V mags show a mean standard deviation of
$\sigma(V)=0.025$\,mag from smooth curves through seasonal data. This includes a typical on-source photometric uncertainty of 0.015\,mag and uncertainties from extinction corrections and calibration-star observations. For each Wing-filter magnitude we assume a linear relation with 
$V$\,mag,  and from the scatter in the $V$ versus Wing magnitudes we infer nightly representative standard deviations: 
$\sigma\left(A\right)=0.068$, $\sigma\left(B\right)=0.051$, and $\sigma(C)=0.037$. 

\subsection{TiO Index}

The TiO index was calculated from the standardized  Wing $A$, $B$, and $C$ filter magnitudes 
from Equation~(\ref{eq:tio_index}) \citep{Wing_1992JAVSO..21...42W}:
\begin{equation}
{\rm TiO\> Index} = A - B - \left[0.13\left(B-C\right)\right].
\label{eq:tio_index}
\end{equation}
                         
The near-IR ($B-C$) color term is applied to account for the decrease in the star’s continuum, and thus between the $B$ and $A$ filters. Because the ($B-C$) color typically varies between $-0.05$ and 
$+0.13$, this photometric correction is small.

The TiO absorption bands increase in strength with later spectral type and decreasing $T_{\rm eff}$. The calibration of the Wing TiO index with spectral type and \teffs are given by  \citet{Wasatonic_2015PASP..127.1010W}, and are based on \citet{Levesque_2005ApJ...628..973L}. The spectral types of the Wing calibration stars extend from $\sim$K5 to M7.5~III stars (in which the TiO-index changes from $\sim 0.25$ to 2.0). The \teff-TiO index relation is given, following \citet{Wasatonic_2015PASP..127.1010W}\footnote{  I.e., the \citet{Wasatonic_2015PASP..127.1010W} relation with slightly updated coefficients, see Appendix.}, by
\begin{equation}
T_{\rm eff}^{Was} = 3902.4  - 421.62\times {\rm TiO}  + 63.931 \times {\rm TiO}^2.
\label{eq:tio_teff}
\end{equation}
This empirical calibration is based on evolved stars that are less luminous and less variable than Betelgeuse, but as \citet{MacConnell_1992AJ....104..821M} point out,  the TiO strength is nearly independent of luminosity.  For stars later (cooler) than $\sim$M7, vanadium oxide (VO) bands and other absorption lines start to contaminate the Wing $B$ and $C$ bands, but this does not affect the present observations of Betelgeuse, which display mean spectral types between M2 and M4.

A small calibration was applied to our $C$ filter (10240\AA) to transform it to the Wing standard system where the original $C$ filter is centered on 10400\AA.  Details of the $C$ band 10240\AA{} color correction and conversion to 
Wing 10400\AA{} magnitudes are given by \citet{Wasatonic_2015PASP..127.1010W}.
As discussed by them, the $C$ bandpass is centered on a 
spectral region with no strong absorption lines and also is near the maximum spectral energy 
distribution of early M-stars like Betelgeuse. Because of this, the $C$ magnitude can serve 
as a proxy for the apparent bolometric magnitude ($m_{\rm bol}$). This permits the luminosity to be computed if the distance is known, and if not, the relative change in luminosity
\citep[see][]{Wasatonic_2015PASP..127.1010W}.

The individual uncertainties of the Wing filter magnitudes propagate into the TiO index through the use of Equation~(\ref{eq:tio_teff}), and give $\sigma\left(TiO\right)=0.089$, and $\sigma\left(T_{\rm eff}\right) \simeq 29$\,K per night. These relatively large individual nightly uncertainties appear to arise from the small aperture of the telescopes, relative high airmass of the stars, and corresponding large extinction corrections. The nightly uncertainties are reduced using weekly mean values (seven-day bins), which mostly include between 1 and 4 nights. The large number of observations clearly reveals the strong trends 
in the photometry.

\subsection{Synthetic Wing Three-filter and $V$-band Photometry}

With the improvements in atomic and molecular data and stellar photospheric modeling, synthetic photometry is a powerful technique to explore the observational consequences of changes in the properties of a stellar atmosphere.  Here we are exploring the differential photometric behavior of Betelgeuse
where all the atmospheric and spectral parameters are kept constant except for \teff.  We generated a grid of spherical MARCS photospheric models \citep{MARCS_2008A&A...486..951G} with  $\log{g_*}   = 0.0$,  $M_\ast  = 15 M_\odot$,  $5$\,\kmss microturbulence, and with
$T_{\rm eff}\left(K\right)=[3200, 3300, 3400, 3450, 3500, 3550, 3600, 3650, 3700]$. 
This \teffs sampling captures the smooth changes in synthetic magnitudes. 
These models are one-dimensional, assume hydrostatic equlibrium, are steady-state, and are computed in local thermodynamic equilibrium. The MARCS-model photospheres have 153,910 opacity samples between  910\AA{} and 20\,$\mu$m.
These photospheres were used to create synthetic spectra for the $V$ band, and the Wing $A$, $B$, and $C$ bands,
using a wavelength sampling of $1.5$\,\kms.
For the carbon, nitrogen, and oxygen (CNO) abundances we adopt the values from \citet{Ryde_2006ApJ...637.1040R}, who scaled the CNO values from \citet{Lambert_1984ApJ...284..223L} (who adopted $T_{\rm eff}=3800$\,K,) to $T_{\rm eff}=3600$\,K. Specifically we adopt
$A_{\rm C} = 8.29$, $A_{\rm N} = 8.37$, and $A_{\rm O} = 8.52$, for the models and the synthetic 
spectra.\footnote{The abundance $A_{\rm X}$ is given through the following definition: $A_{\rm X}\equiv\log{n_{\rm X}} - \log{n_{\rm H}} + 12$, where $n_{\rm X}$ is the  number density.}
We adopt the carbon isotopic ratio $^{12}$C/$^{13}$C$=7$ from \citet{Harris_Lambert_1984ApJ...281..739H}. The linelists used include  metals and CH, ${\rm C}_2$, CN, NH, OH, MgH, AlH, SiH, CaH, CrH, TiO, ZrO, VO, H$_2$O, and FeH  \citep[see][and references therein]{MARCS_2008A&A...486..951G}.

The synthetic spectra were then reddened with the same interstellar parameters adopted by \citet{Levesque_2005ApJ...628..973L}, namely $A_V=0.62$, $R_V=3.1$, with the reddening prescription of \citet{CCM1989ApJ...345..245C}.  Following \citet{Levesque_2005ApJ...628..973L}, this reddening  also includes any potential contribution from the amorphous silicates in the low optical depth circumstellar envelope
\citep[see e.g.,][]{Rowan_1986MNRAS.222..273R}.
We adopt a constant photospheric angular diameter of 44\,mas \citep[see for example,][]{Montarges_2016A&A...588A.130M}.
Synthetic Wing three-filter and $V$-band photometry was performed using the filter transmissions and detector response of the Optec SSP-3. The zero-point  magnitude  offsets were derived from the Vega model spectrum of \citet{Vega_1994A&A...281..817C}, using the calibrations given by
 \citet[][Table 2]{Wing_1992JAVSO..21...42W}.  Table~\ref{tab:sensitivity} gives the resultant
 synthetic reddened magnitudes for MARCS-model spectra with different effective temperatures.

\begin{table}
{  
\begin{center}
\caption{Temperature Sensitivity of {\rm Betelgeuse} Synthetic Photometry to $T_{\rm eff}$}
\label{tab:sensitivity}
\begin{tabular}{lccc}
\tableline\tableline
Band     &  $C_0$  &  $C_1$   & $C_2$   \\  \tableline
$V$           &  142.05 & $-7.414\times 10^{-2}$    &  $+9.683\times 10^{-6}$ \\ 
$\sigma\left(C_V\right)$   & $3.84$ & $0.223\times 10^{-2}$ & $0.322\times 10^{-6}$\\
$A$           &  116.62 &  $-6.134 \times 10^{-2}$  &  $+7.899\times 10^{-6}$ \\ 
$\sigma\left(C_A\right)$   & $3.87$ & $0.224\times 10^{-2}$ & $0.325\times 10^{-6}$\\
$B$          &   28.81 &   $-1.576\times 10^{-2}$ & $+1.96 \times 10^{-6}$ \\
$\sigma\left(C_B\right)$   & $2.82$ & $0.164\times 10^{-2}$ & $0.237\times 10^{-6}$\\
$C$          &   2.46   &  $-1.421\times 10^{-3}$     & \nodata \\
$\sigma\left(C_C\right)$   & $0.08$ & $0.023\times 10^{-3}$ & \nodata \\
\tableline
\end{tabular}
\tablecomments{Reddened magnitude $ = C_0 + C_1 T_{\rm eff}+ C_2 T_{\rm eff}^2$}
\end{center}
}
\end{table}

These dedicated Betelgeuse models also permit the determination of a model-dependent 
\teff--TiO calibration. {  We find that for a given TiO index the inferred \teffs is between
10\,K (near 3300\,K) and 70\,K (near 3600\,K) lower than that given by Equation~(\ref{eq:tio_teff})}. There are several possible causes for this small difference aside from different stellar parameters and abundances, e.g., \citet{vanBelle_2009MNRAS.394.1925V} find empirically a 
non-negligible  variation in \teffs within an individual RSG spectral-type class, namely $\pm \ge 100$\,K. 
It is also known that 1-D photospheric models are not perfect representations of real (physical) M supergiants,
especially in the outer layers where strong molecular lines are formed \citep[see, for example][]{Arroyo_Torres_2015A&A...575A..50A, Chiavassa_2011A&A...535A..22C}. 
Equation~(\ref{eq:tio_teff})
is also based on less luminous cool evolved stars. Also, \citet{Davies_2013ApJ...767....3D} have pointed out that \teffs based on TiO band-head fits to 1D MARCS models are systematically lower, by over 100\,K,  than those based on spectral energy distributions. 

In \S4 we will use the MARCS synthetic photometry to model the photometric behavior of Betelgeuse when the surface is covered with regions with different  values of \teff, and then compare these to the observed photometry.

\section{Results\label{section:results}}

\begin{figure}[t]
\begin{center}
\epsscale{0.85}
\epsfig{file=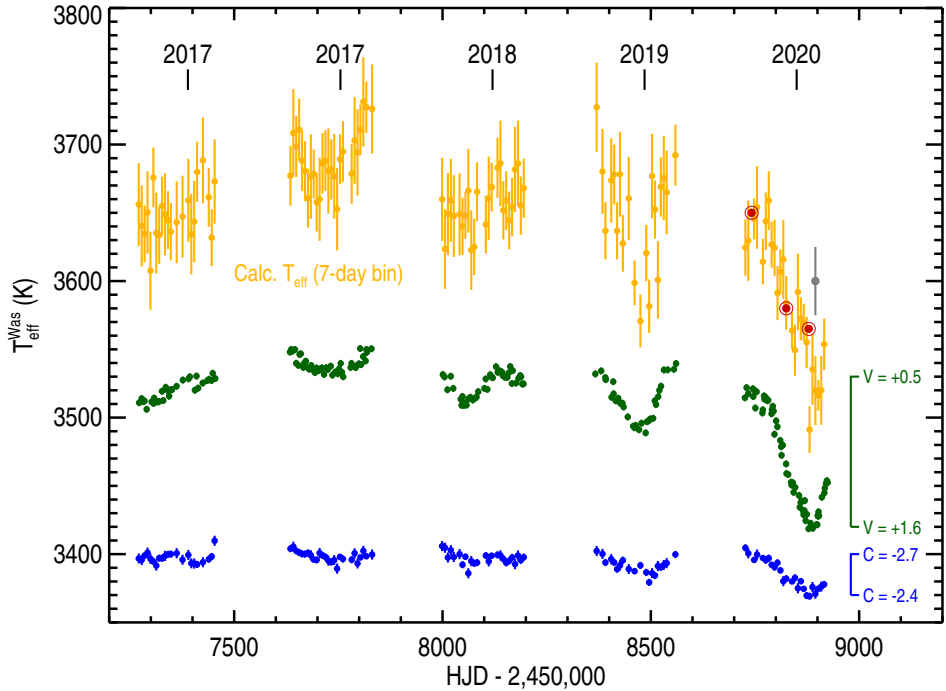, angle=90., width=13.5cm}
\caption{Five years of photometry, described in \S2.1, and the calculated $T^{Was}_{\rm eff}$ (gold) derived  from the TiO index using Equation~(\ref{eq:tio_teff}). The \teffs and $C$ magnitudes have been averaged into 7 day bins.  Also shown are the $V$\,mag (green, unbinned) and $C$\,mag (blue) data.  The $1\sigma$ error bars are also shown, but for $V$ they are similar to the symbol size.  The  {  three red bulls-eye symbols are the TiO-based \teffs previously reported {  in}  \citet{Guinan_2019ATel13341....1G} and \citet{Guinan_2020ATel13439....1G}, and that occur on
2019 September 15, December 7, and 2020 January 31, or HJD-2450000=8742, 8825, 8895, respectively.}
The gray circle and error bar are the mean \teff derived from the optical spectrum of
\citet{Levesque_2020ApJ...891L..37L}.  The TiO-based \teff{}s are significantly cooler than derived from the optical spectrum. The TiO-based \teff{}s are clearly correlated with  $V$ over the past 5 seasons, including during the recent large dimming event. The correlation with $C$ is also present, but the magnitude range is smaller.}
\label{fig:tio_teff}
\end{center}
\end{figure}

Figure~\ref{fig:tio_teff}  shows the \teffs averaged on a grid of 7 day bins for the past 5 years, with the propagated uncertainties reflecting the number of nights included in each bin.  During the Great Dimming the discrepancy between the TiO-based \teffs  and that based on the strength of TiO band heads in the 4000--6800\AA\ spectral region by \citet{Levesque_2020ApJ...891L..37L} is immediately apparent. In the most recent observing cycle, the TiO-based \teffs declined from $T_{\rm eff}\simeq 3645\pm 15$\,K {  (2019 September 21/HJD=2458748; 7 day bin)} to $\simeq 3520 \pm 25$\,K {  (for individual 2020 February 15/HJD=2458895, 22, and 29 7 day bins)}, i.e., $\Delta T_{\rm eff}\simeq -125$\,K. (For this discussion we  omit the 7-day bin at 
$3491\pm 17$\,K). The difference between September and February of $\Delta T_{\rm eff}\simeq -125$\,K  corresponds to a change of spectral type from M2 to M4 in six months. \citet{WW1978ApJ...222..209W} noted that changes of spectral subtype by 0.5  in RSGs seldom occurred over such a short interval.  Figure~\ref{fig:tio_teff}  also shows the clear correlation between $V$ and TiO-based \teffs over the last 5 years of photometry, including the large dimming event where no sudden change in trend is apparent as $V$  increased beyond its typical range. This correlation can be understood as a result of the strong temperature sensitivity of TiO molecular absorption; lower gas temperatures lead to increased TiO formation and deeper absorption. During the recent dimming event $V$ increased by about $+1.1$~mag from 2019 September to 2020 mid-February (a reduction in flux by a factor of 2.75 in flux). The $C$~mag, which is a measure of the $1.02 \mu$m continuum flux, also follows the same \teffs trend, but it only increased by  
$\Delta C \simeq +0.25$~mag (a reduction in flux by a factor of 1.25), the largest on record.

The most interesting result  from the TiO photometry is the significant difference in \teffs at minimum $V$-band brightness, i.e., $\le 3520$\,K and  the  $T_{\rm eff}=3600 \pm 25$\,K derived by \citet{Levesque_2020ApJ...891L..37L} from the strength of TiO band heads between 4500 and 6800\AA. 
Given that both temperature estimates are based on TiO bands from low-lying energy levels it is unlikely that such a difference would occur because of different formation heights during a dynamic event.  If the star had uniformly reduced its \teffs from $3650$\, to $3600$\,K while maintaining the same angular size,  then $V$ would increase by $+0.2$ and the  flux in the $V$ band would only decrease by $\simeq 17$\%.  If the \teffs had decreased to $3520$\,K  $V$ would increase by 0.6 and the flux decrease by 43\%. However, the observed increase in $V$ was $+1.1$.

\begin{figure}[t]
\begin{center}
\epsscale{0.85}
\epsfig{file=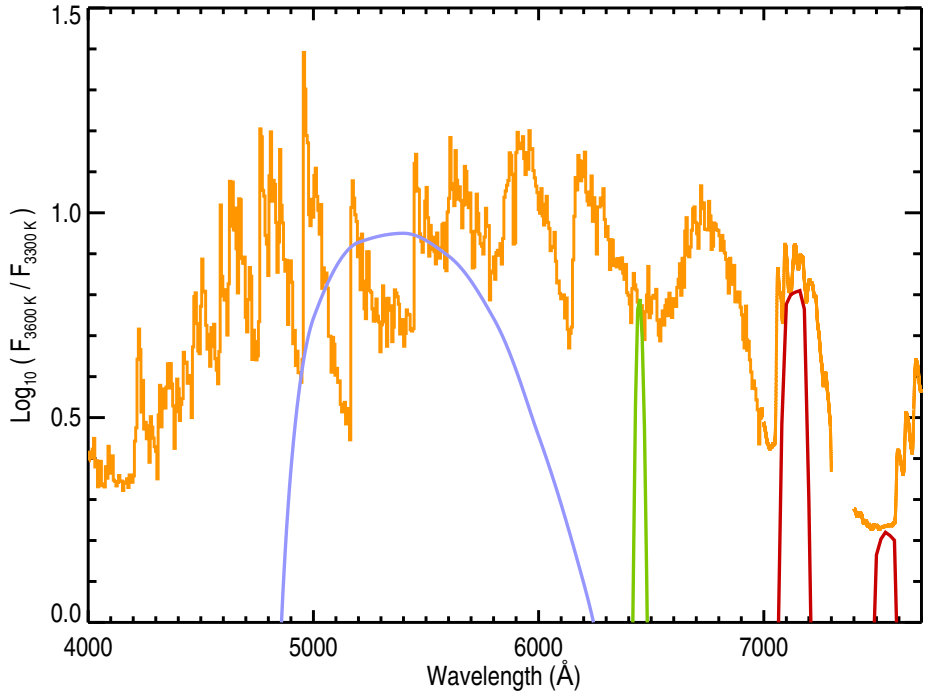, angle=90., width=13.5cm}
\caption{The logarithm of the flux density ratio of $T_{\rm eff}=3600$\,K to $T_{\rm eff}=3300$\,K spectra computed from the MARCS models with all other stellar parameters kept constant. The spectral ratio has been binned to $\sim 6$\AA.   Relative filter responses have also been plotted with arbitrary offsets for clarity.
The blue curve is the $V$-band response, the red curves are the Wing $A$ and $B$ bands, and the green curve the VLT-SPHERE H$\alpha$-continuum filter. As can be seen, the flux for $T_{\rm eff}=3600$\,K ranges between  a factor of 3 to 12 brighter within the $V$ band, with a filter
mean of $\sim 7$.  The flux ratio is $\sim 8$ in the Wing $A$ filter, while in the continuum-dominated spectral region of the  Wing $B$ filter at 7540\AA, the ratio is only $\sim 1.7$ (0.22 dex).} Importantly, below 4500\AA{} the flux ratio declines from the high values in $V$ band.
\label{fig:show_ratio}
\end{center}
\end{figure}

Clues to the origin of these apparent discrepancies are the two VLT-SPHERE H$\alpha$-continuum images ($\lambda$6449\AA, $\Delta\lambda=41$\AA) obtained in 2019 January and December (Montarg\`es et al. 2020). These reveal that in December
the southern hemisphere was dimmer than the north, showing brightness variations across the star. Once it is recognized that regions of different \teffs may exist across the visible hemisphere, the different empirical constraints can be reconciled.

Synthetic photometry shows that the $V$ magnitude is, as expected, very sensitive to $T_{\rm eff}$ (see Table~\ref{tab:sensitivity}), and by illustration Figure~\ref{fig:show_ratio} shows the flux ratio of the MARCS spectrum for $T_{\rm eff}=3600$\,K to that for $T_{\rm eff}=3300$\,K. The continuum flux ratio in the Wing $B$ filter (and $C$, not shown) is $\sim 1.7$, being much less sensitive to $T_{\rm eff}$ than the TiO-dominated $V$ and Wing $A$ (7190\AA) filter bands, the latter includes the strong  $\gamma(0,0)$ band. Here the mean flux ratios are $\sim 7-8$.

Consider the case of a photospheric region covering an area of $50\%$ of the star's visible hemisphere with $T^{cool}_{\rm eff} \sim 3300$\,K and the rest of the visible hemisphere having $T_{\rm eff}=3600$\,K. The cool component would have very low emission across most of the $V$- and $A$-band spectral region, as shown in Figure~\ref{fig:show_ratio}, where the TiO bands dominate. The resulting combined spectrum (3600 and 3300\,K) would then yield  a $T_{\rm eff}^{optical} \le 3600$\,K,  but geometrically diluted, i.e, the total flux is reduced by the fraction of the star covered in cooler material.   The TiO index, on the other hand, is determined primarily by the $A$ and $B$ filters, which will have different relative contributions to the combined spectrum. The flux in $A$ band, from the cool component, like $V$, would contribute very little to the total spectrum. However, $B$ will contribute much more flux than $A$ because it is much less sensitive to \teff. The combined spectrum, would appear to have a  deeper $A$-filter absorption than the 3600\,K component alone, lowering the $T_{eff}$ from 3600\,K to a value between that of the two components, which would explain the 
TiO-based $T_{\rm eff} \simeq 3520$\,K. In this scenario $T_{\rm eff}^{cool} < T_{\rm eff}^{TiO} <
T_{\rm eff}^{optical} \le T(3600{\rm K})$.

 In this example the $V$ brightness would be $\simeq 55\%$ of that without the cool component. During the Great Dimming  the reduction in $V$ brightness  was greater than this, which is easily accommodated with a fractional area of the cool component of $>50$\%.

\begin{figure}[t]
\begin{center}
\epsscale{0.85}
\epsfig{file=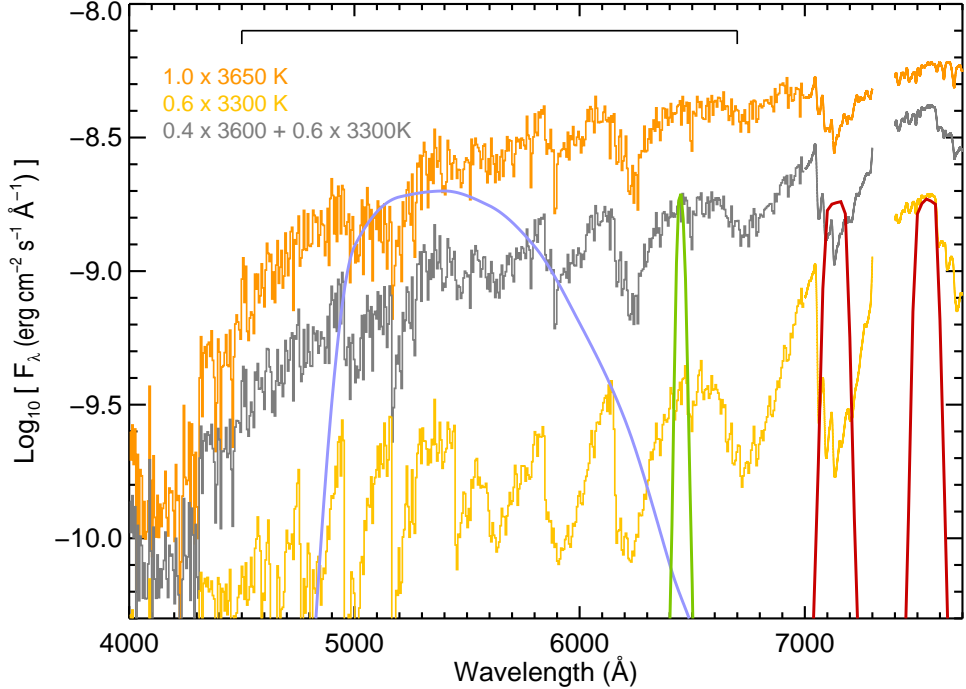, angle=90., width=13.5cm}
\caption{A two-component photospheric model illustrating the dimmed state on 2020 February 15. The top orange curve is for $T_{\rm eff}=3650$\,K representative of its more typical state, and the gray curve represents 60\% of the visible hemisphere covered by $T_{\rm eff}=3300$ and 40\%  covered with
$T_{\rm eff}=3600$\,K. For clarity relative filter responses have also been plotted with arbitrary offsets.
The blue curve is the $V$-band response, the red curves are the Wing $A$ and $B$ bands, and the green curve the VLT-SPHERE H$\alpha$-continuum filter. 
The 4500- 6700\AA\ spectral region (indicated by black bar at top) has been binned to match  
\citet[][Figure~1]{Levesque_2020ApJ...891L..37L}.  The two-component gray curve is a factor of $\sim 2$ (0.3 dex) less than the 3650\,K curve, a similar difference to the 2004 and 2020 observations of 
\citet{Levesque_2020ApJ...891L..37L}.
The gold lower curve shows the cool component's contribution, which illustrates how the $V$-band contribution is
almost negligible, and why the combined spectrum in this region resembles a geometrically diluted $T_{\rm eff}=3600$\,K spectrum. However, below 4500\AA, the cooler spectrum is not reduced as much, and the differences become less significant.}
\label{fig:two}
\end{center}
\end{figure}

For this geometric dilution interpretation the difference between the normal- and dimmed-state spectra would be mostly independent of wavelength between 4500 and 6700\AA. However, below 4500\AA\   the ratio of hot- to cool-component fluxes decreases, and the differences 
between the combined spectra at these wavelengths also decreases, as pointed out by \citet{JCMT_2020ApJ...897L...9D}.  This model would also explain the much smaller decrease in the combined continuum flux at $C$ (1.02\,$\mu$m).
The concept of a mean $T_{\rm eff}$ is no longer useful in this case. 
Figure~\ref{fig:two} shows a simple two-component model that illustrates these points: the cool 
component has $T_{\rm eff}=3300$\,K but with a  fractional area filling factor of 60\%. This model
shows the spectra in the same fashion as \citet[][Figure~1]{Levesque_2020ApJ...891L..37L}
to aid comparison.  Between 4700 and 6500\AA\ the overall flux decrease between the 3650\,K and two-component model matches the difference ($\simeq 0.3$\,dex) between the \citet{Levesque_2020ApJ...891L..37L} 2004 and 2020 spectra. Furthermore, the
combined spectrum is not that different from a diluted $T_{\rm eff}\simeq 3600$\,K
which may explain the \teffs found by \citet{Levesque_2020ApJ...891L..37L}.

Different combinations  of \teffs and area fraction are possible; the cooler the
cool component, the closer the total combined spectrum resembles that of
a diluted $T_{\rm eff}=3600$\,K spectrum. The
10240\AA\ region (not shown) is reduced by only 0.1\,dex (or $\Delta C=+0.25$~mag) as observed.
Furthermore, since the $C$ mag is a measure of the luminosity of the visible hemisphere,  and $L \propto T_{\rm eff}^4$, then for the model above the relative luminosity $L(2-{\rm comp})/L({\rm 3650\>K}) \simeq 0.4 + 0.6(3300/3650)^4 \simeq 0.80$, which matches $\Delta C=+0.25$ and 
 $L(2-{\rm comp})/L({\rm 3650\>K}) \simeq  2.512^{-0.25} \simeq 0.79$, as expected.

We can exploit the Wing three-filter and $V$-band photometry to further examine the temporal behavior during the last two seasons including the dimming episode. We construct a simple model where the star is composed of two different components: one with a typical $T_{\rm eff}=3650$\,K  with an area fraction $(1-A)$, where $A$ is the variable fractional area of the visible hemisphere, and one component with $T_{\rm var}$ and area $A$. This extra component may mimic hot spots, or cool regions, as observed by previous interferometry studies. The total flux from the system is then
\begin{equation}
F_{\rm total} = \left(1-A\right) F_{\rm 3650\,K} + AF_{T_{\rm var}}.
\label{eq:combo}
\end{equation}

This model assumes that the angular size of the photosphere is constant and uses steady-state MARCS models. It also assumes that each component includes both limb and disk emission contributions, i.e., each component is sector-shaped. Using the $T_{\rm eff}$-magnitude relations given in Table~\ref{tab:sensitivity}, we compute the apparent magnitudes of the 
$A_{\rm mod}$, $B_{\rm mod}$, $C_{\rm mod}$, and  $V_{\rm mod}$ magnitudes for a fine two-dimensional grid of $A$ ($0\to1$) and $T_{\rm var}$ ($3000\to 3750$\,K) and then find the least-squares minimum for each 7 day bin, namely we minimize
\begin{equation}
  \left( {V_{\rm obs}-V_{\rm mod}\over{\sigma\left(V\right)}} \right)^2 +
  \left({A_{\rm obs}-A_{\rm mod}\over{\sigma\left(A\right)}}\right)^2 +
  \left({B_{\rm obs}-B_{\rm mod}\over{\sigma\left(B\right)}}\right)^2 +
    \left({C_{\rm obs}-C_{\rm mod}\over{\sigma\left(C\right)}}\right)^2.
\end{equation}
The $\sigma$s are the individual uncertainties for each 7 day bin, and because the $V$ magnitudes have the smallest uncertainties they have the greatest weight in the solution. The TiO-dominated $A$ and $V$ magnitudes are strongly correlated, and the continuum-dominated $B$ and $C$ magnitudes are  also correlated.

\begin{figure}[t]
\begin{center}
\epsscale{0.85}
\epsfig{file=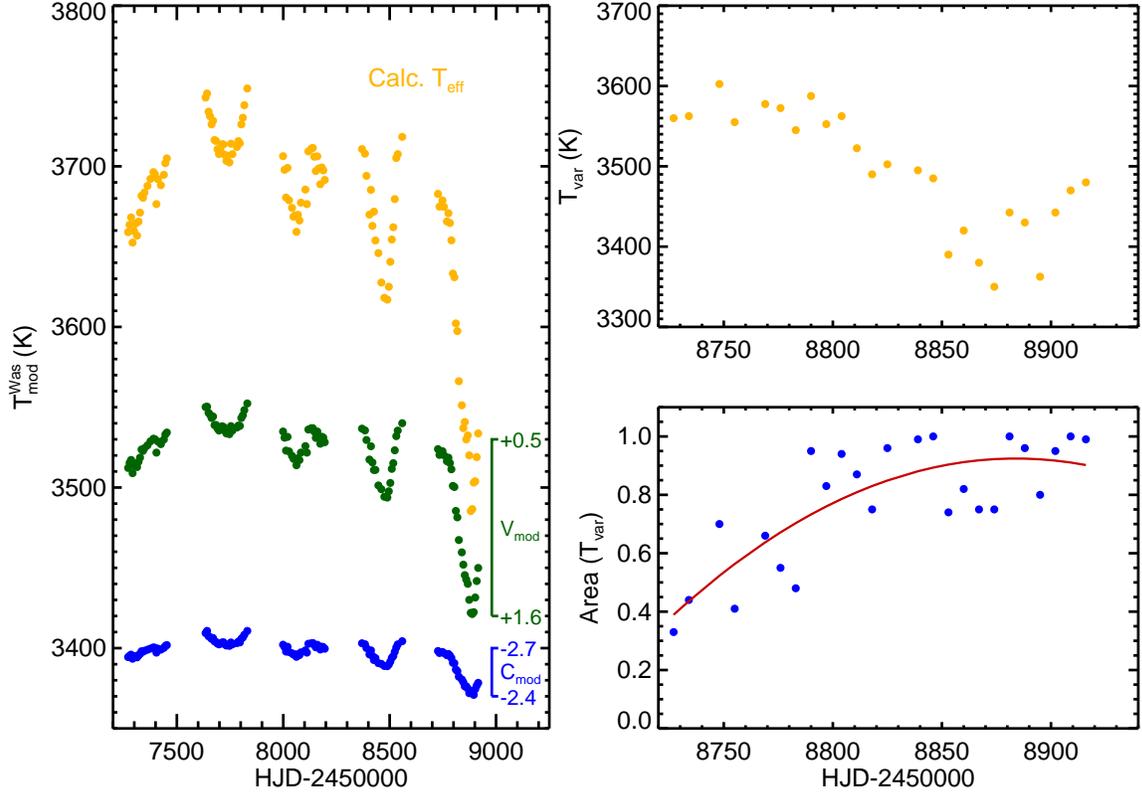, angle=90., width=14.5cm}
\caption{Results of the two-component model, Equation~(\ref{eq:combo}), for each 7 day bin. Left: the model \teffs for the whole star,
$T_{\rm mod}^{\rm Was}$, calculated from Equations~(\ref{eq:tio_index}) and (\ref{eq:tio_teff}) with the Wing and $V$ magnitudes from each best-fit solution. The corresponding $V$ and $C$ magnitudes are shown below.  Note that if the internal TiO-\teffs calibration from the MARCS models had been used the computed  \teffs would be lower by 10--70\,K (see \S2.3). Top right Top: the values of $T_{\rm var}$ corresponding to the best-fit solutions for the season of the Great Dimming.  Bottom right: the corresponding area fractions of the photosphere covered with $T_{\rm var}$, with a simple polynomial trend curve (red). Both parameters of the two-component model show significant scatter, but the general trends of decreasing \teffs and increasing $A$ are seen with increasing date, until the deep $V$-band minimum when $T_{\rm var}$ begins to increase. }
\label{fig:two_comp}
\end{center}
\end{figure}

The best-fit solutions for the magnitudes were also converted to $T^{\rm Was}_{\rm mod}$ using Equations~(\ref{eq:tio_index}) and (\ref{eq:tio_teff}), and the results for the dimming cycle are shown in Figure~\ref{fig:two_comp}.
(Recall that $T_{\rm var}$ is the \teffs of the input MARCS models.)
There is a tension between $T_{\rm var}$ and $A$, for example a reduction in stellar flux can be caused by a reduction in \teffs for a given fractional area, or from a smaller reduction in $T_{\rm var}$ but with a great area fraction.  There is considerable scatter in the solutions which is not real. 
Significant changes
in the brightness and/or the areas of large features in hydrodynamic simulations occur on timescales greater than 7 days \citep{Chiavassa_2009A&A...506.1351C}. 
The scatter probably reflects the simple, i.e., nonphysical nature of the model, and noise in the observational data. The assumption that each temperature component be sector-shaped is unlikely to be good, especially when a convective feature emerges within the stellar disk. 

During the 2018/2019 season (not shown), $T_{\rm var}$ has a high area factor and starts near $3650$\,K, declining to $\simeq 3550$\,K by mid-season and then increasing again to $\sim 3650$\,K.
  Figure~\ref{fig:two_comp} shows the 2019/2020 season, and September starts with $T_{\rm var}\simeq 3550-3600$\,K with $A\sim 0.3 - 0.4$, slightly cooler than
the typical \teff,  but  during the Great Dimming 
$T_{\rm var}$ steadily declines while $A$ increases. By 2019 December 28 the model suggests
$T_{\rm var}\sim 3485^{+15}_{-165}$\,K, with $A\sim 0.85^{+0.15}_{-0.25}$, cooling further to 
$T_{\rm var}=3360^{+65}_{-45}$\,K  on 2020 February 15 with a similar  $A$-range.
During the subsequent  increase in $V$ brightness in 2020 March, $T_{\rm var}$ increased while the fractional area remained the same. The VLT-SPHERE 2019 December image suggests an area factor on the lower end of that suggested by the model. Because the photometry is spatially unresolved this model does not tell us where the cool region is or whether it is comprised of multiple cool components. This very simple model is probably the best that can be justified with the available four photometric bands. The two-component example shown in Figure~\ref{fig:two} included 60\% of the area covered by $T_{\rm eff}=3300$\,K, although material at $3200$\,K leading to more distinct
band heads could not be ruled out. Clearly, in reality the star probably has a distribution of $T_{\rm eff}\left(A\right)$.  The addition of multi-epoch optical spectra to the analysis during this dimming event would improve the constraints on the geometric dilution term $A$, and hence $T_{\rm var}$.

\section{Discussion}

We have found that the Wing three-filter near-IR and $V$-band photometry can be explained by an inhomogeneous photosphere, with one component having a large fractional area that
is significantly cooler than the typical $T_{\rm eff}=3650$\,K expected for an M2~Iab RSG. 
The \teffs of cool component is less than the mean \teffs based on the TiO index, i.e., $< 3520$\,K.    \citet{JCMT_2020ApJ...897L...9D} have also reached a similar conclusion based on their observations of reduced sub-mm fluxes\footnote{  Given that the sub-mm emission is from an optically thick region above the
photosphere \citep{Eamon_2017A&A...602L..10O}, the 20\% sub-mm flux reduction implies a $\sim 20$\% reduction in local gas temperature if the angular size of the emitting region is unchanged, larger than 
the 5\% adopted in that paper.} during the dimming event, and an examination of model multi-component optical spectra. The appearance of new dust is not required to interpret the Great Dimming.

\citet{Levesque_2020ApJ...891L..37L} derived a $T_{\rm eff}=3600\pm 25$\,K for the minimum of the Great Dimming, which is only slightly cooler than expected for a M2~Iab spectral type. To explain the factor of two flux reduction (0.3~dex) observed between their 2004 and 2020 spectra they proposed that the dimming was
predominantly caused by attenuation from new dust in the line of sight. The grains were assumed to be large because of the relatively wavelength-independent flux reduction between 4500 and 6700\AA.
However, if large dust grains were injected into our line of sight then the Wing $A$ (7190\AA) and $B$ (7540\AA) spectral regions would be similarly affected, and the TiO-based \teffs values would be expected to be near $3600$\,K, which they are not. 

If dust did attenuate the $V$-band flux then some thermal dust-emission signatures might be expected.  However, \citet{Gehrz_2020ATel13518....1G}  reported infrared photometry at 1.2, 2.2, 3.6, 4.9, 7.9, and 8.8\,$\mu$m, that does not reveal evidence of an increase in emission compared to previous observations reported 50 years prior \citep{Gehrz_1971ApJ...165..285G}. The 2020 values are consistent within the error bars longward of 4.8\,$\mu$m with those reported in 1971, and they are 0.2--0.3~mag fainter than those obtained between 1991 April 25 and May 3 \citep{Fouque_1992A&AS...93..151F} when the nearest $V$ measurement epoch, 3 to 4 weeks prior,  had $V\sim 0.55$ \citep{Kri_1992IBVS.3728....1K}. The decrease in  $J$ band brightness is similar to what we find for the $C$ band (1.024\,$\mu$m).   \citet{JCMT_2020ApJ...897L...9D} presented a schematic dust model to mimic the dimming event, adopting alumina and Mg-Fe silicate dust grains.
The addition of new dust leads to an increase in flux longward of 1.25\,$\mu$m rising to 60\% at 8\,$\mu$m. Because these simulations are idealized (e.g., a spherical shell rather than a localized dust cloud that might reflect starlight into a different direction) the infrared evidence cannot conclusively rule out the formation of any new dust; however, it does not provide support for the presence of dust to attenuate the $V$-band fluxes.

From SOFIA-EXES spectra, \citet{Harper_2020ApJ...893L..23H} found that the ratio of circumstellar emission-line fluxes of
[\ion{Fe}{2}] at 25.99\,$\mu$m  and [{\ion{S}{1}] at 25.25\,$\mu$m to the local continuum fluxes was not significantly different between the dimming episode and two previous epochs when the star was in its brighter normal state. The last two observations were made with the same aperture. The continuum captured in the EXES apertures is mostly stellar with a contribution from extended circumstellar silicate dust emission \citep{Harper_2001ApJ...551.1073H}.
 Given that the collisionally excited emission lines and the photospheric and dust continuum are formed by different physical processes in different spatial regions, the constancy of the EXES 
 line-to-continuum flux ratios suggests that the stellar continuum near 26$\mu$m has not
significantly changed. It would be unlikely that the changes in continuum and CSE emission would change in the same direction by the same amount. The \citet{JCMT_2020ApJ...897L...9D}  schematic dust model predicts a continuum increase by a factor of 1.87 at 26\,$\mu$m (T. Dharmawardena 2020, private communication) which is not evident in the SOFIA-EXES spectra.

The observed enhancement in \ion{Mg}{2} h \& k emission and NUV continuum in 2019 October, 
 prior to the  Great Dimming, could have resulted in an ejection of hot dense plasma potentially leading to dust formation \citep{Dupree_2020ApJ...899...68D}. However, the dust (if formed) may not necessarily be in our line of sight. 

While some dust-related event may have been associated with the Great Dimming, the evidence suggests that the dimming in $V$ (as well as  in the Wing three-filter TiO near-IR photometry) was due to a significant fraction of the visible stellar surface having a \teffs $\le 250$\,K cooler than normal.

\subsection{Polarimetry: Changes in Illumination and/or Dust?}

There have been reports of polarimetry obtained at epochs near the Great Dimming, and changes
in dust have been described as a potential cause of the changes in polarization
 \citep{2020RNAAS...4...39C, 2020RNAAS...4...47C,Safonov_2020arXiv200505215S}. Here we briefly discuss this point.

The observations of Betelgeuse by \citet{Hayes_1984ApJS...55..179H} found that both the degree of linear polarization and position
angle in the $B$ band changed significantly, and in an orderly fashion, during four consecutive observing seasons. The non-repeating polarimetric pattern had a timescale of $\sim$ 0.5--1\,year, and it was suggested that the origin may be related to the rotation of large convective cells. This time-scale is consistent with convective (but non-giant) cells \citep{Schwarzschild_1975ApJ...195..137S}, and also for $\sim 0.1 R_\ast$ cells in hydrodynamic simulations that have time-scales of a few months to one year \citep{Chiavassa_2009A&A...506.1351C}.  Linear continuum polarization may occur from atmospheric Rayleigh scattering and/or Mie scattering off circumstellar dust. The observed dependence of linear polarization with wavelength suggests atmospheric  Rayleigh scattering of radiation from a non-axisymmetric illumination source at, or near, the stellar surface. \citet{Auriere_2016A&A...591A.119A} have confirmed this interpretation with observations of Betelgeuse's line de-polarization spectrum, which cannot be created with dust scattering 
alone.\footnote{The polarized spectrum from the photosphere may subsequently be polarized by circumstellar dust scattering.}

\citet{2020RNAAS...4...39C, 2020RNAAS...4...47C} found that
the 4250\AA\ and 4800\AA\ polarization decreased during the $V$ brightness minimum, which could
result from a decrease in asymmetric illumination. However, \citet{Safonov_2020arXiv200505215S} employing differential speckle polarimetry found
that the net polarization brightness did not change until 2020 February, when it subsequently increased as the $V$ brightness increased. They suggest the initial constancy supports the idea of a dust cloud along the line of sight because the extended envelope would still be illuminated by the rest of the star.  However, that does not easily explain the subsequent increase in polarization brightness that occurred when the star returned towards its normal brightness.  Dust is known to exist close to the photosphere, \citet{Kervella_2016A&A...585A..28K} detected dust at $\sim 3R_\ast$, and \citet{Haubois_2019A&A...628A.101H} detected dust near  $\sim 1.5R_\ast$. Changes in polarization during the dimming may instead be related to changes in illumination from the stellar surface and/or changes in the distribution of existing dust.

\subsection{Radial Velocity - Light Variations}

\citet{Dupree_2020ApJ...899...68D}  presented photospheric radial velocities from the last two short-period ($\sim 430$ days) cycles obtained with the STELLA robotic observatory \citep{Strassmeier_2004AN....325..527S}.  Figure~\ref{fig:vrad} reproduces these data\footnote{Using WebPlotDigitizer: https://automeris.io/WebPlotDigitizer}  together with our $V$ magnitudes.  

During these cycles the maximum (positive) radial velocity (i.e., the maximum contraction velocity of the photosphere) occurs about 35 days ($\simeq 0.08$\,Period (P)) after the time when the star is faintest and coolest 
(see Figures~\ref{fig:tio_teff} and \ref{fig:vrad}).  In addition, the maximum stellar expansion velocity occurs during the star's, only partially covered, broad maximum brightness. 
This phasing is similar to the behavior of pulsating yellow supergiant Cepheid variables. For example, in the case of the well observed classical Cepheid prototype, $\delta$~Cep (F5~Ib -- G1~Ib; $P = 5.366$ days), minimum brightness of the star also occurs $\sim$0.085\,P prior to the star's maximum contraction radial velocity 
\citep[see][and references therein]{Engle_2017ApJ...838...67E}. This behavior indicates that the observed $V$, temperature, and radial-velocity variations of Betelgeuse (observed over the star’s dominant $\sim$430 day period) may arise from radial pulsations. But unlike Cepheids, which have repeatable light and radial-velocity variations, Betelgeuse has multiple periods and complex, semi-regular light variations.

It is interesting that the {\em range} of radial-velocity variation ($\Delta v=10$\,\kms) during the dimming cycle was higher than the previous cycle ($\Delta v=6$\,\kms), and the former is typical of the larger variations in radial velocity  previously recorded  \citep[][and references therein]{Goldberg_1984PASP...96..366G}, 
suggesting a particularly dynamic photospheric event. This could arise from a more vigorous pulsation episode or a larger than usual giant convection cell, or possibly both.  The coherent nature of the last two seasons of photospheric radial velocities and the mean \teff, and the coincidence of the minimum of the short- and long-period variations suggests that the 
Great Dimming was a stronger pulsation or shock event, leading
to enhanced photospheric cooling. During the dimming event the measured photospheric radial velocities are weighted toward the hotter diminishing photospheric component, and therefore because of the smooth changes in radial velocity with time, the larger cooling component presumably shares similar infall velocities.

\begin{figure}[t]
\begin{center}
\epsscale{0.85}
\epsfig{file=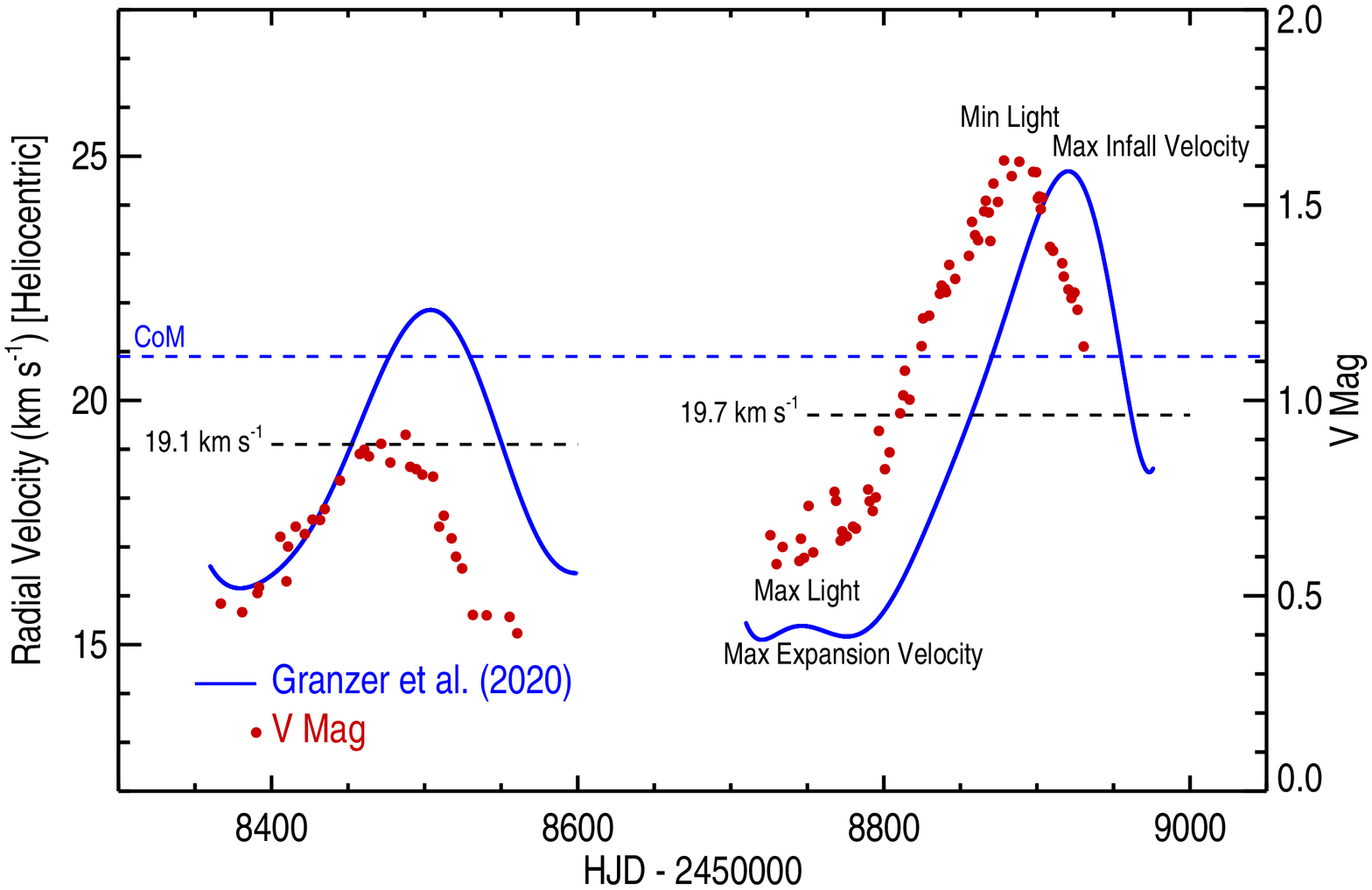, angle=90., width=13.5cm}
\caption{Radial velocities as reported by \citet{Dupree_2020ApJ...899...68D} fit with a smooth curve and our
$V$ magnitudes. For ease of comparison the $V$\,mags are plotted with  largest values (minimum brightness) at the top. The center-of-mass (CoM) stellar radial velocity of $20.9 \pm 0.3 (1\sigma)$\,\kmss is taken from \citet{Harper_2017ApJ...836...22H}, is based on a combination of estimators, and is consistent with the ALMA 
spatially-resolved molecular diagnostics \citep{Kervella_2018AA...609A..67K}.  During these epochs the mean photospheric radial velocity is smaller than the CoM velocity. The mean values for each cycle are shown (average of leading minimum and maximum).  The offsets in mean radial velocities with the CoM velocity could be caused by the radial velocity contribution arising from the long 
($\sim 5.6$ -- 6.0 yr) period. \citet[][and references therein]{Sanford_1933ApJ....77..110S} find peak-to-peak radial velocity ranges of between 4.1 and 6.1\,\kms. }
\label{fig:vrad}
\end{center}
\end{figure}

Previous observations show similar, but not identical, behavior. \citet{Dupree_2013EAS....60...77D} have discussed the relationship between radial velocity {  and $V$-band observations}
for a 16 year span, including the radial-velocity measurements of \citet{Myron_1989AJ.....98.2233S}.
{  Even though the coverage is typically not uniform, they find} that there are times when the opposite correlation of maximum radial velocity and maximum brightness (hence temperature) is observed, while at another time the opposite occurs.
 \citet{Gray_2008AJ....135.1450G} presented radial-velocity and temperature measurements from the neutral lines of  \ion{V}{1} at 6251.83\AA, \ion{Fe}{1} at 6252.57\AA, and \ion{Ti}{1} at 6261.11\AA, between epochs 2002.5 and 2007.5. These data suggest a general pattern between \teffs-$v_{rad}$, where the temperature increased before the expansion phase {  (blue shifted lines)}, and cooling occurred before the contraction phase, similar to that shown in Figure~\ref{fig:vrad}. However, unlike the 
 \citet{Gray_2008AJ....135.1450G} results, during the last two seasons
(with more dense coverage) Betelgeuse shows  a clear correlation between the size of the ranges of $v_{\rm rad}$ and the \teff.

\citet{Kravchecnko_2019A&A...632A..28K} have studied the temporal relation of $v_{\rm rad}$ and \teffs
in the more luminous RSG $\mu$~Cep (M2-~Ia; \citep{KM1989ApJS...71..245K}) and in radiation-hydrodynamic (RHD) simulations. Following the ideas of  \citet{Gray_2008AJ....135.1450G}, \citet{Kravchecnko_2019A&A...632A..28K} developed a time-dependent {  scenario} to interpret the quasi-periodic variations as a photospheric shock, driven by activity in the deeper hidden convection zone. However, a noticeable difference in these studies is that the phase relation between $v_{rad}$ 
and $V$ is not as coherent as that shown in 
Figure~\ref{fig:vrad} {  for Betelgeuse}.  We note that sparse temporal coverage makes it hard to draw firm conclusions from previous observational studies, and also existing RHD models for RSGs do not show such coherent variations. 

Perhaps the key point to take away from the  high precision STELLA radial-velocity data and our photometry is that the correlation seen in the
last two cycles between $V$ and $v_{\rm rad}$, namely the phase offset and amplitudes,  shows no sudden deviation or 
discontinuity. This suggests that a new phenomenon has not occurred, e.g., dust formation in our line of sight that changes the  $V$ brightness discontinuously during the dimming. The $V$ and radial velocities measurements, along with Wing photometry, are closely correlated in the last two seasons where we have observations to show it.

\section{Conclusions}

The Wing three-filter and $V$-band photometry indicate a significantly cooler mean $T_{\rm eff}$ during the 2019/2020 Great Dimming (see Figure~\ref{fig:tio_teff}) than that inferred from the strength of the TiO band heads in the 4000--6800\AA\ spectra by \citet{Levesque_2020ApJ...891L..37L}. These findings can most simply be reconciled by the presence of a large area, $\ge 50$\%, of even cooler photospheric material. The observed $V$-band spectral region is then a geometrically diluted spectrum of the normal $T_{\rm eff}$ component. This picture also explains why the differences in the 2004 and 2020 spectra below 4500\AA{} are smaller than
in the  rest of the spectral region, as pointed out by \citet{JCMT_2020ApJ...897L...9D}. This is a consequence of the \teff-sensitivity of $V$-band spectral region, as shown in 
Figure~\ref{fig:show_ratio}. No new dust is required to explain these observations.  Independent information from observations made during the Great Dimming, namely IR photometry  \citep{Gehrz_2020ATel13518....1G} and SOFIA-EXES spectra  \citep{Harper_2020ApJ...893L..23H} also do not reveal any signature of dust emission.  Changes in linear polarization are not so easily interpreted, but they may be related to changes in photospheric illumination
of pre-existing dust, known to be present close to the photosphere.

The photometric data support the hypothesis that the  Great Dimming was {  a greatly enhanced} (amplified) continuation of regular changes in mean \teff, $V$, and radial velocity, over the 
short 430-day period cycle. The deep minimum occurred close to the time predicted.  The increased 
range in radial velocity during the dimming cycle, see Figure~\ref{fig:vrad}, may reflect a dynamic origin of the cooling of large areas of the photosphere.

It is hoped that by combining all the observations obtained before, during, and after the Great Dimming  a more complete understanding of RSG variability can be achieved.

\appendix

Differential photometry was carried out in the usual pattern of 
{\it sky-comparison-variable-comparison-(check star)-sky} that generally repeated 3-4 times 
for each filter. Typical integration times per measurement were 30-50 s.. These 
individual measures were averaged to form nightly means. The star was generally observed 
for 50-60 min per night. The primary comparison star was 
$\phi^2$~Ori (40 Ori, G9~III-IV; $V =+4.09$\,mag, $B-V = 0.96$). $\phi^2$~Ori is an older 
inactive high-velocity star; no significant variability ($<3$\,mmag) is indicated from Hipparcos photometry \citep{ESA_1997ESASP1200.....E}. The check star, 
$\delta$~Ori, (Bellatrix, HD 35468, B2~V, $V=+1.64$, $B-V= -0.22$.) was observed differentially 
with respect to the comparison star on most nights in the V band. Bellatrix does seem to 
show possible small light variations  ($\pm 0.015$\,mag). Wing standard stars were also observed 
and used as alternate check stars. Differential atmospheric extinction corrections were applied, 
and the mid-times of the observation converted to Heliocentric Julian Dates.
Equation (\ref{eq:tio_index}) is based on a larger set of calibration observations than used in \citet{Wasatonic_2015PASP..127.1010W}.  All calibration observations obtained between 1996 and 2009 were used to revise the atmospheric extinction and Wing-system transformation coefficients.
Equation (\ref{eq:tio_teff}) differs from the 2015 version by less than 15\,K  between 3200 and 3800\,K.

\bibliography{harper_TiO_2020Oct12}

\acknowledgments

Financial support for G.M.H. was provided by NASA through a SOFIA Betelgeuse Flash Archival Research award \#07-0073 issued by USRA. N.R. acknowledges support from the Royal Physiographic Society in Lund through the Stiftelse Walter Gyllenbergs fund and M\"arta and Erik Holmbergs donation. 
The Wasatonic Observatory (Allentown, PA) is supported by Villanova University.
We thank T.  Dharmawardena for providing numerical details of their dust model, and we also thank RWH for invaluable assistance. We acknowledge the reviewers for their helpful comments which improved the presentation of this paper. This research has made very extensive use of NASA's Astrophysics Data System Bibliographic Services. This research has also made use of the VizieR catalogue access tool, CDS, Strasbourg, France  \citep{Vizier_2000A&AS..143...23O},  including the 5th Edition of the Catalog of Infrared Observations \citep{Gezari_1999yCat.2225....0G}.

\vspace{5mm}
\facilities{(Wasatonic Observatory)}

\end{document}